\documentclass[aps,prl,twocolumn,showpacs,groupedaddress]{revtex4-1}
\usepackage{amsmath,amssymb,subfigure,wasysym}
\usepackage[dvipdfmx]{graphicx}
\DeclareMathAlphabet{\mathpzc}{OT1}{pzc}{m}{it}

\begin{document}

\title{Topological lasing and self-induced transparency in two level systems}

\author{Laura Pilozzi}\email{Corresponding author: laura.pilozzi@isc.cnr.it}
\author{Claudio Conti}
\affiliation{Institute for Complex Systems, National Research Council (ISC-CNR), Via dei Taurini 19, 00185 Rome, Italy}


\begin{abstract}
The use of virtually lossless topologically isolated edge states may lead to a novel class of thresholdless lasers operating 
without inversion. One needs however to understand if topological states may be coupled to external radiation, and act as 
active cavities. We study a two-level topological insulator and show that self-induced transparency pulses can directly
excite edge states. We simulate laser emission by a suitable designed topological cavity, and show that it can emit tunable radiation.
For a configuration of sites following the off-diagonal Aubry-Andr$\acute{e}$-Harper model\cite{Aubry,Harper}, we solve the Maxwell-Bloch equations in the time domain and provide a first principle confirmation of topological lasers. Our results open the road to a new class of light emitters 
with topological protection for applications ranging from low-cost energetically-effective integrated lasers sources, also including silicon photonics, to strong coupling devices for studying ultrafast quantum processes with engineered vacuum.
\end{abstract}

\pacs{42.65.Sf,42.50.Md,02.70.Bf}

\maketitle

\textit{Introduction ---}   In the context of transport phenomena, two or three-dimensional Bloch and Anderson models, paradigmatic for periodic and disordered structures, allow to observe a crossover from extended to localized states\cite{anderson} at a critical degree of disorder \cite{ishii,Thouless}.
However, a new class of structures, the topological insulators, show a localization phase transition\cite{Verbin} in one-dimension (1D).
Originally described in the tight-binding formulation for electrons\cite{Aubry}, and recently for the study of localization properties of acoustic\cite{Yang}, electromagnetic\cite{Wang,Rechtsman,Lu,Kraus,ganeshan} and matter waves\cite{roati}, topological insulators are characterized by the presence of peculiar edge states, corresponding to a conducting surface for a bulk insulating material.
The geometric phase of the bulk crystal determines the existence of these edge states and, correspondingly, they are protected, i.e. stable against any perturbation.

Recently the study of localization properties in topological systems has been extended to the class of resonant photonic crystals\cite{poddubny1,pod} sustaining topologically protected boundary states \cite{Posha}, also involving the exciton-photon coupling\cite{Karzig}.
The possibility of topologically protected states in resonant systems opens the challenge of realizing topologically sustained lasers,
i.e, lasers based on edge states. These devices are expected to benefit of the intrinsic isolation, and hence may eventually operate at very low threshold, or without population inversion. Indeed the potential absence of loss reduces virtually to zero the gain needed for the laser operation.

In these terms, the first question to consider is if resonant topologically isolated systems can be directly excited from external inputs, and seemingly if topologically isolated states can emit coherent light into propagating modes when acting in a laser device.

In this Letter we show that a direct excitation of topological edge states is achievable in chains of two-level systems (TLS) by the use of an ultrashort self-induced trasparency (SIT) pulse.
Our results are based on the simulation of the Maxwell-Bloch equations \cite{Taflove,Ziolkowski} and we study SIT in a resonant topological insulator (RTI) where index modulation is given by either the TLS, or by the background dielectric function.  In analogy with the disordered case\cite{Folli}, the spatial distribution of the active layers localize the SIT pulse that would otherwise induce a travelling population inversion. This localization is a fingerprint for edge states detection and sustain tunable laser emission.
\begin{figure}[b]
\subfigure{\includegraphics[width=0.43\columnwidth]{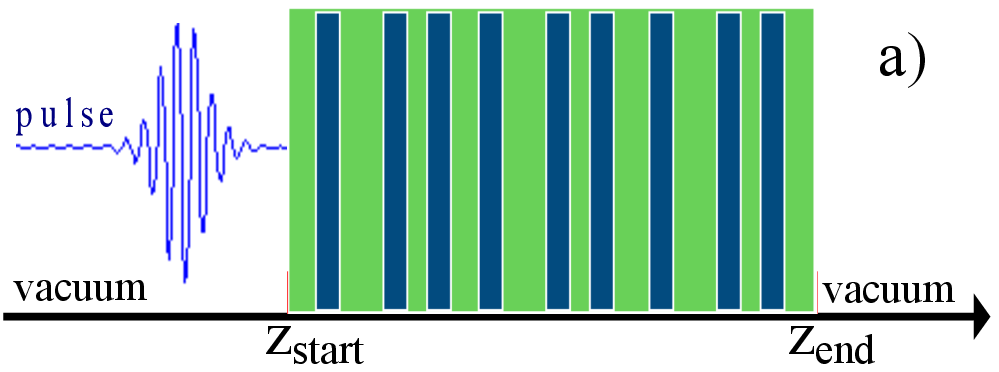}}\quad
\subfigure{\includegraphics[width=0.53\columnwidth]{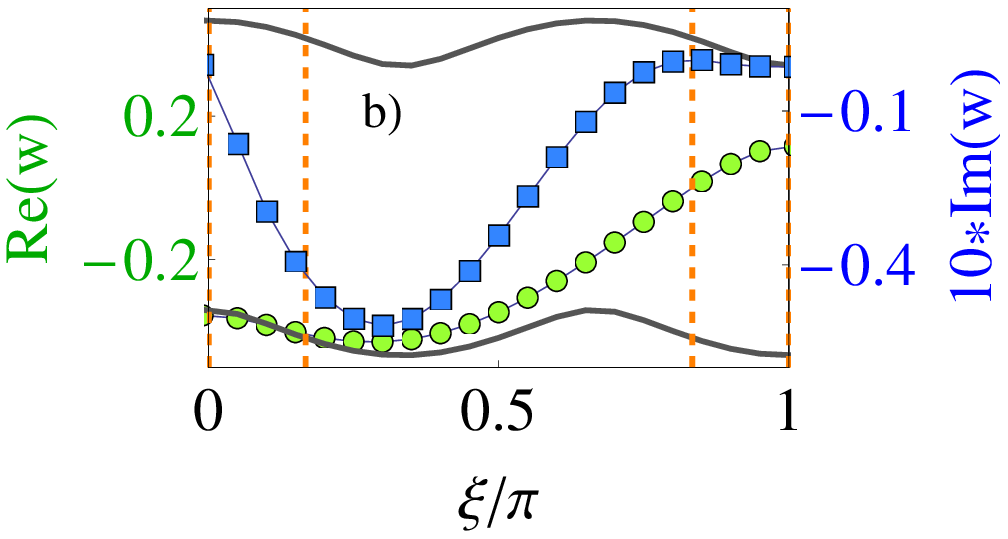}}
\caption{(Color online) a) Sketch of the 1D chain of resonant two-level layers (blue) in an homogeneus bulk (green) of frequency-independent dielectric function $\varepsilon _b$. b) Real (green circles) and imaginary (blue squares) part of the left edge states frequency for $\beta=1/3$, $\eta  = 0.2/\pi$, $\omega _o  = 1.533\,eV$, $\varepsilon _b  = 12.25$. The dashed vertical lines show the $\xi$ values ($\xi/\pi=0,1/6,5/6,1$) where, for simmetry reasons, edge states do not exist. The continuos lines mark the gap boundaries.\label{fig:f12}}
\end{figure}

\textit{Structure and edge states dispersion ---}    A schematic of the system considered is shown in Fig.~\ref{fig:f12}a). The structured region consists of resonant two-level layers A (width $L_A$) in an homogeneus bulk of frequency-independent dielectric function $\varepsilon _b$. Two configurations will be considered. In a uniform structure (US), the resonant layers, with their background dielectric function $\varepsilon _a=\varepsilon _b$, are arranged in a sequence with centers in $z_n  = d_o\left[ {n + \eta d_n^H } \right]$, where $d_n^H = \cos (2\pi\beta n + \phi )$ is the Harper modulation\cite{Harper}.
In a Bragg structure (BS), the resonant layers have $\varepsilon _a\neq\varepsilon _b$ and the widths $b_n$ of the dielectric layers B are modulated as: $b_n  =b_o\left[ {1 + \eta d_n^H } \right]$.

These distributions define a 1D bichromatic periodic lattice (period $d_o$) modulated by a secondary lattice with strength $\eta$.
The phase shift $\phi$ governs the localization phase transition and 
the modulation frequency $2\pi \beta$ determines the number of topological boundary states in the gap.
The structure is periodic with $\beta^{-1}$ resonant layers in the unit cell and period $d =\beta^{-1} d_o$.

\textit{Uniform structure ---} We choose $d_o = \lambda _o /2$ with $\lambda _o  = 2\pi c /(\omega _o\sqrt {\varepsilon _b })$, in order to center 
the photonic band gap of the ordered stack ($\eta=0$) at $\omega_o$  (TLS resonance).
The A layers have radiative (non-radiative) $\Gamma _o$ ($\Gamma$) decay rate, dielectric constant $\epsilon_a$ and reflection coefficient $r_A (\omega ) =  -i/(w + i)$, with $w=(\omega  - \omega _o  + i\Gamma)/\Gamma _o $, with a local Lorentz-like dispersion:
\begin{equation}
\chi _A (\omega ) =  -  \frac{{\hbar^2 c^2 }}{\omega_o^2} \frac{{L_A q^3 }}{16 \pi}\frac{{(q^2L_A^2  - 4\pi ^2 )^2 }}{{16\pi ^4 \sin ^2 (qL_A /2)}}\frac{1}{w}
\end{equation}
with $q=\omega\sqrt{\epsilon_a} /c$.
The poles of the reflection coefficient of the whole structure give the left-edge state frequencies $w_\ell$, solutions with negative imaginary part\cite{Posha} of:
\begin{equation}
e^{2iqs_1 } + (w - i)^2 e^{2iq(s_1  + s_2 )} + (w + i)^2 + (w^2 + 1)e^{2iqs_2 }  = 0\text{.}\\
\end{equation}
The symmetry $w_r(\mp\xi)=w_\ell(\pm\xi)$, with $\xi =\phi - \pi/6$, gives the right-edge modes $w_r$.
The states lay within the gap centered at $\omega _o$ with bounds given by $Tr(T)=\pm 2$, where $T$ is the single period transfer matrix.

Figure~\ref{fig:f12}b) shows the $\xi$ dependence of the real part of the left-edge state frequency. 
When $\xi$ varies in $(0,\pi)$ the edge modes traverse the band gap, bounded by the straight lines; 
the imaginary part $\Im[(\omega-\omega_o+i\Gamma)/\Gamma_o]$ gives their inverse lifetime.

\begin{figure}[t]
\includegraphics[width=0.8\columnwidth]{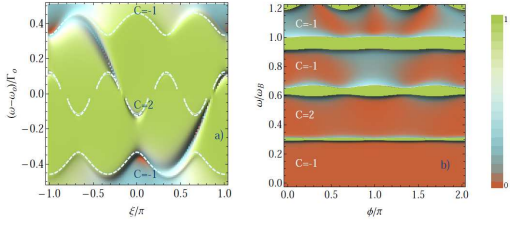}
\caption{Squared amplitude of the reflection coefficient ${r_\infty (\phi , \omega )}$ from the left side of the semi-infinite US chain a) and BS chain b) with Chern numbers C.\label{fig:f2}}
\end{figure}

Figure~\ref{fig:f2}a) shows the reflectivity $\left| {r_\infty (\xi , \omega )} \right|^2$  for $\Gamma_o  = 10^{-2} \omega_o$ and $\Gamma  = 10^{-2} \Gamma_o$ and the corresponding Chern numbers C\cite{PoshaAr}. The edge states correspond to dips in $\left|r_\infty\right|^2$: their experimental observation requires fine spectral resolution and a high ratio between the radiative and non-radiative decay rates $\Gamma_o/\Gamma$. In these terms, 1D systems with weak losses and a large resonance strengh $\Gamma_o/\omega_o$ are ideal candidates for edge-state detection.

\textit{Bragg structure ---}  For different dielectric constants of layers A and B the spectral gaps of the structure with $\eta=0$ at 
integer multiples of $\omega _B  =\pi c/(L_a \sqrt {\varepsilon _a }  + b_0 \sqrt {\varepsilon _b } ) $ split in $\beta^{-1}$ gaps.
We choose $\varepsilon _a= 1$ and $\varepsilon _b= 12.25$, $b_0  = 200\;nm$ and $L_a = 48\;nm$, so that $\omega _B  = 0.828\;eV$.

\begin{figure}[b]
\subfigure{\includegraphics[width=0.49\columnwidth]{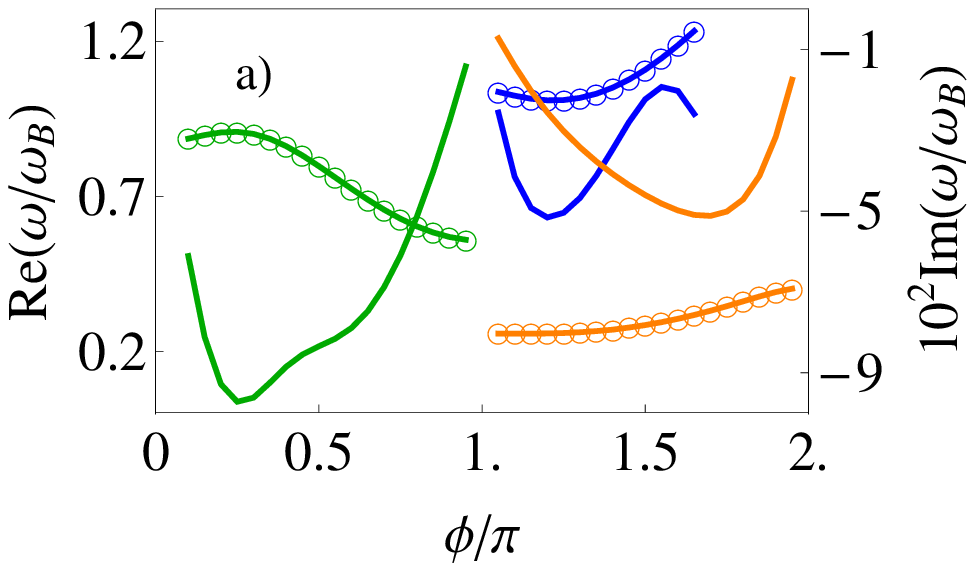}}
\subfigure{\includegraphics[width=0.49\columnwidth]{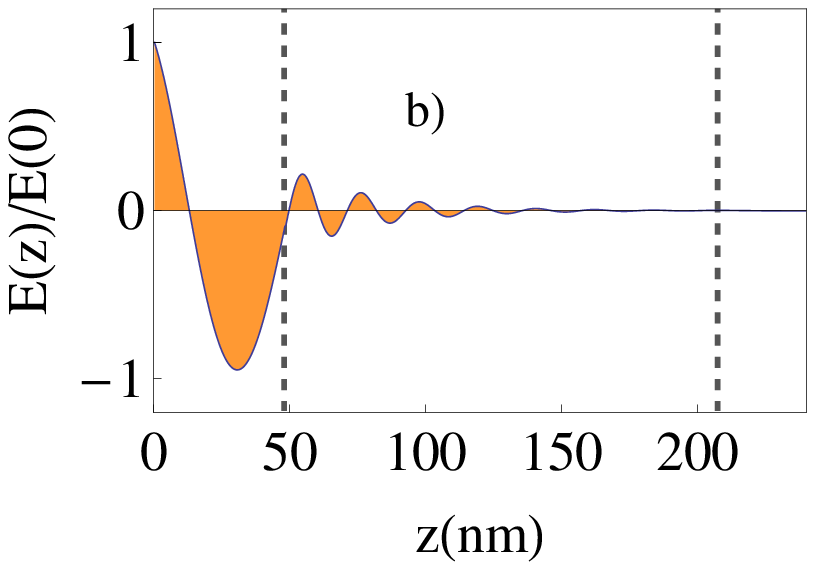}}
\caption{(Color online) a) real (open circles) and imaginary (straight line) part of the left edge states frequency for $\beta=1/3$, 
$\eta  = 0.5$, $\varepsilon _a  = 1$, $\varepsilon _b  = 12.25$. b) field intensity distribution inside the system for the configuration with $\phi=0.7\pi$.
\label{fig:f3}}
\end{figure}

For $\beta=1/3$ and $\eta=0.5$, Fig.~\ref{fig:f2}b) shows the reflection coefficient $|r_\infty (\phi , \omega )|^2$ from the left-edge of the semi-infinite system. Figure~\ref{fig:f3}a) shows the real and imaginary part of the left-edge eigenfrequencies, and the field intensity distribution (b) for $\phi=0.7\pi$, with the localized mode profile at $\omega=\omega_{BS}(0.7\pi)$.

\textit{Time-domain dynamics ---}  To obtain the electric field amplitude, polarization, and population inversion we describe the dynamics of light propagation in the RTI trough the Maxwell-Bloch equations:
\[
\begin{array}{l}
 \mu _0 \partial _t H_y  =  - \partial _z E_x  \\ 
 \varepsilon _0 \partial _t E_x  =  - \partial _z H_y  - \partial _t P_x  \\ 
 \end{array}
\]
with $P_x = 2\gamma N\rho _1 $, where N is the resonant dipole density and $\gamma$ is the dipole coupling coefficient, and
\begin{equation}
\partial _t \left[ {\begin{array}{*{20}c}
   {\rho _1 }  \\
   {\rho _2 }  \\
  {\delta\rho _3 }  \\
\end{array}} \right] =- \left( {\begin{array}{*{20}c}
   {\gamma_2 } & {-\omega _o } & 0  \\
   {  \omega _o } & {\gamma_2 } & {-2\omega _R }  \\
   0 & {  2\omega _R } & {\gamma_1 }  \\
\end{array}} \right)\left[ {\begin{array}{*{20}c}
   {\rho _1 }  \\
   {\rho _2 }  \\
   {\delta\rho _3 }  \\
\end{array}} \right] + \left[ {\begin{array}{*{20}c}
   0  \\
   { 2\omega _R }\rho _{30}  \\
   0  \\
\end{array}} \right]
\end{equation}
where ${\delta\rho _3 }={\rho _3 }-{\rho _{30} }$, the vector $\left[ {\begin{array}{*{20}c}
   {\rho _1 } & {\rho _2 } & {\rho _3 }  \\
\end{array}} \right]^T $
is the state density vector, with $\rho _1$  ($\rho _2$) proportional to the in-phase (in-quadrature)
polarization, $\rho _3$ proportional to the inversion population and
$\omega _R  = \gamma E_x/\hbar$ the Rabi frequency; $\gamma _1$ and $\gamma _2$ denote the population and polarization relaxation rates while $\rho _{30}$ is the initial population inversion.
The corresponding susceptibility, $\chi (\omega ) =  - N\gamma \rho _1/(\varepsilon _o E_x)$ is:
\begin{equation}
\chi (\omega ) =  \frac{2N\gamma^2}{{\varepsilon _o}\hbar } \frac{{\omega _o }}{ {(i\omega  + \gamma _2 )^2  + \omega _o^2 } }
\end{equation}

\textit{SIT pulse in topological insulators ---} Following Ref.\cite{Ziolkowski} we consider the evolution of a pulse that coming from vacuum ($\epsilon_o$) moves in the structured region of Fig.~\ref{fig:f12}a) with an initial sech profile: $E_x (0,t) = E_o \text{sech}\left[ {10(t - \tau /2)/(\tau /2)} \right]\sin \left[ {2\pi f_o t} \right]$. We choose the pulse frequency resonant with the medium, $2\pi f_o=\omega_o/\hbar$, the pulse duration $ \tau  = 191\;fs$ and $E_o$ to have a $2\pi$ pulse\cite{McCall} after the reflection on the input face due to the $\epsilon_o, \epsilon_b$ mismatch.
The one-dimensional periodic active medium consists of $N_c$ cells with resonant layers $L_A$ wide, separated by slices of trasparent material with relative permittivity $\epsilon_b$ = 12.25. The dielectric layers where the TLS are not present have $\rho_{30}$ =0 and widths $s_n=z_{n+1}-z_{n}-L_A$ for the US and $s_n=b_n$ for the BS.
We model the US system as a collection of two level atoms with density N=$10^{24}m^{-3}$ and dipole coupling coefficient $\gamma=1.4 \times 10^{-27}Cm$ such that $N\gamma^2=w \Gamma_o \chi_A(\omega_o)$. Moreover we fix $\gamma _1= \gamma _2=$0.23 THz. For the BS structure we choose N=$10^{24}m^{-3}$ and $\gamma=1\times 10^{-29}Cm$.

\begin{figure}[b]
\includegraphics[width=0.9\columnwidth]{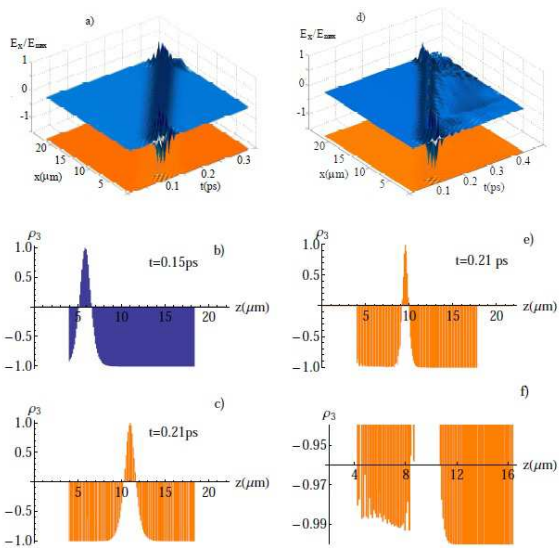}
\caption{Field and population inversion $\rho _3$(z, t) spatial profile for different observation times $t_i$ for the US structure with $\xi_{TC}=0.5\pi$.
a-c) Reference configuration with $\gamma=1\times 10^{-29}Cm$. d-f) Topological configuration with $\gamma=1.4\times 10^{-27}Cm$ \label{fig:f6} }
\end{figure}

\begin{figure}[t]
\subfigure{\includegraphics[width=0.48\columnwidth]{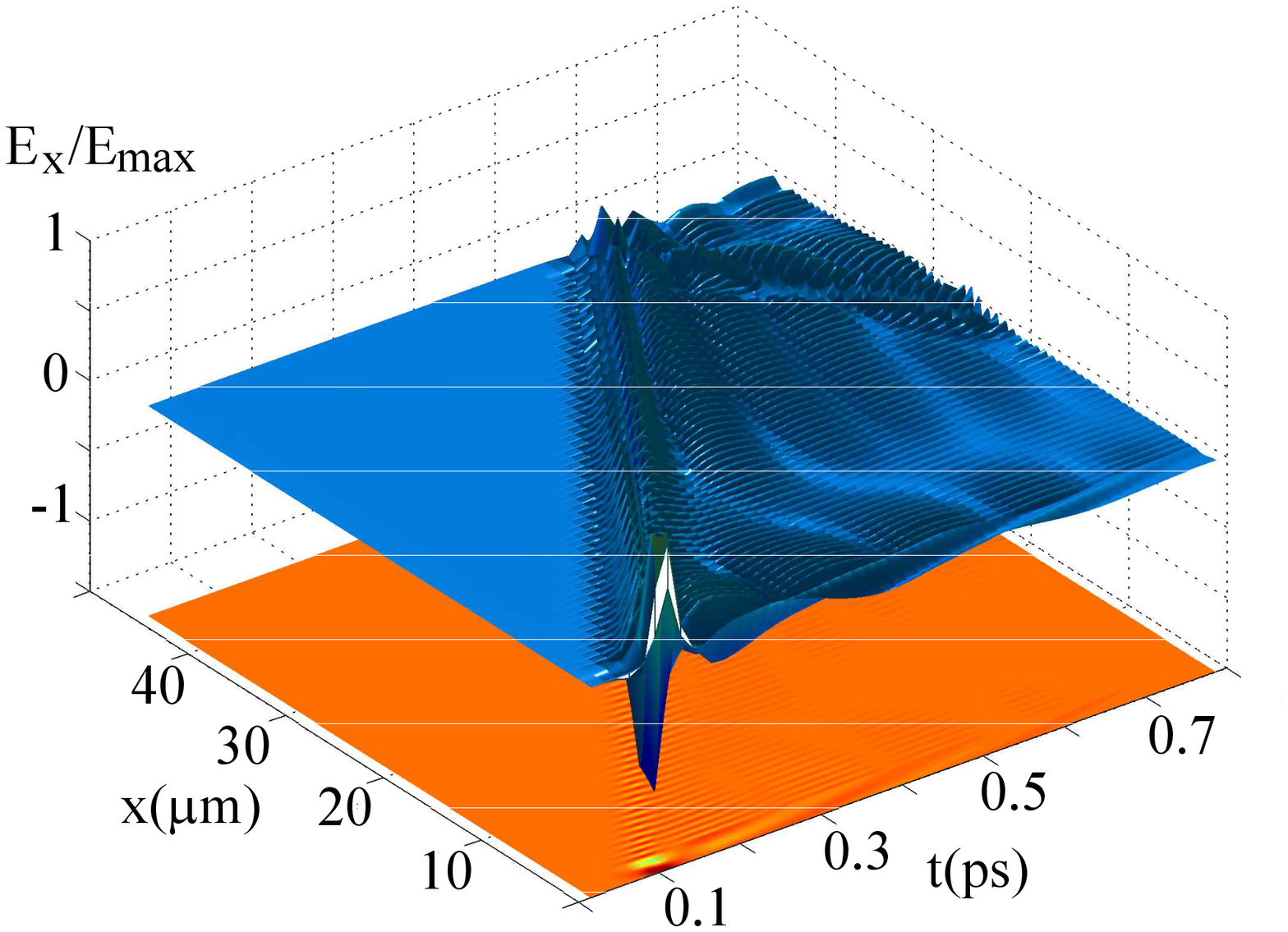}}\quad
\subfigure{\includegraphics[width=0.48\columnwidth]{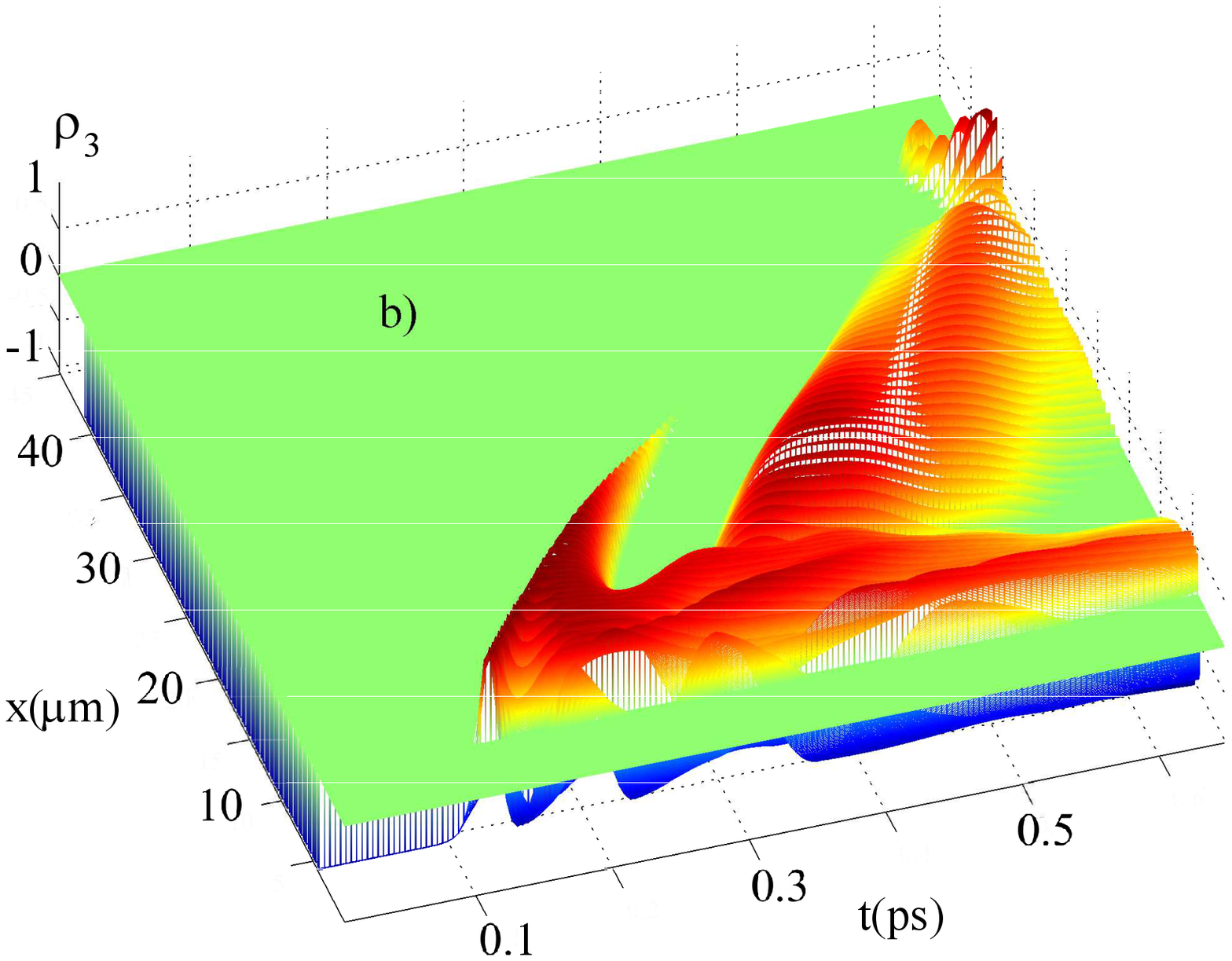}}
\caption{Field a) and population inversion b) $\rho _3$(x, t) spatial profile for different observation times $t_i$ for the BS structure with $\phi_{TC}=0.7\pi$. \label{fig:f7} }
\end{figure}

We analize the field and population inversion spatial profile for different observation times $t_i$.
The $E_x(z,t)$ and $\rho _3$(z, t) plots are shown in Fig.~\ref{fig:f6} for the US chain for which $N_c=40$ and in Fig.~\ref{fig:f7} for the BS one for which $N_c=50$.
For both the structures $z_{start}=4\mu m$ and a $2\mu m$ layer of material $\epsilon_b$ is present at the front and rear side.
To point out that the edge states are excited by the external input, we compare a topological configuration (TC) with a reference one (RC). In particular, according to the dispersion relations (Fig.~\ref{fig:f12}b),~\ref{fig:f3}b)), we choose:

- for the US, $\xi_{TC}=0.5\pi$, with an edge state at the frequency $\nu_\ell =369.4$ THz and lifetime $\tau_\ell =0.98 ps$;

- for the BS, $\phi_{TC}=0.7\pi$ with an edge state at the frequency $\nu_\ell =121.48$ THz and lifetime $\tau_\ell =0.014 ps$.

The reference configuration is given by the choice $\epsilon_a=\epsilon_b$ for the BS chain. In the uniform one we switch off the edge state by simply decreasing the dipole coupling coefficient.
We remark that in the US the modulation in the refractive index, given by the pulse interaction with matter, is the origin of both the gap and the edge state.

In absence of an edge state, Fig.~\ref{fig:f6}a), the incident laser pulse with its initial intensity and width evolve in a steady-state envelope and propagates without attenuation at a constant velocity. As a consequence the maxima $\rho_3(z,t)$ =1, i.e. population inversion, Fig.~\ref{fig:f6}b)and c), track, in space and time, the same path for the excitation through the structure. The BS chain gives similar results.
On the contrary in the topological configuration, the localization at the input face of the spectral component of the pulse corresponding to the edge mode frequency is evident. For the US topological configuration, where the refractive index modulation is given only by the contribution of the resonance, the pulse propagates with a lower dispersion (Fig.~\ref{fig:f6}d) with respect to the BS configuration (Fig.~\ref{fig:f7}a). 
In both the cases, as a consequence of localization, the main pulse no longer meets the SIT condition and undergoes attenuation due to absorbtion by the TLS. The asimmetry in the $\rho_3(z,t)$ shape for the US chain, shown in Fig.~\ref{fig:f6}e), and in an enlarged scale in Fig.~\ref{fig:f6}f), is a fingerprint of this localization.
The $\rho_3(z,t)$ shape in Fig.~\ref{fig:f7}d) for the BS chain shows attenuation of the main peak and evidence for the onset of the edge mode propagation for times longer than its lifetime.

\textit{Topological lasing ---}  Our challenge is to show that the interplay of topological localization and amplification can be exploited to design mirrorless laser systems in analogy with random structures\cite{conti}.
To this end, with the resonant layers as the light-amplifying material, we study edge modes in the stimulated emission process. We start with the two-level system population initially inverted in the upper state  $\rho_{30}$ =1 and add, following Ref.\cite{Slavcheva}, as the only source a stochastic term with Gaussian statistic in the electric field evolution $E_x=\sqrt{-2\xi_Eln(a)}cos(2\pi b)$ with a and b random numbers uniformly distributed in (0,1) interval and variance $\xi_E=10^{-3} V^2m^{-2}$.
\begin{figure*}[t]
\includegraphics[width=1.8\columnwidth]{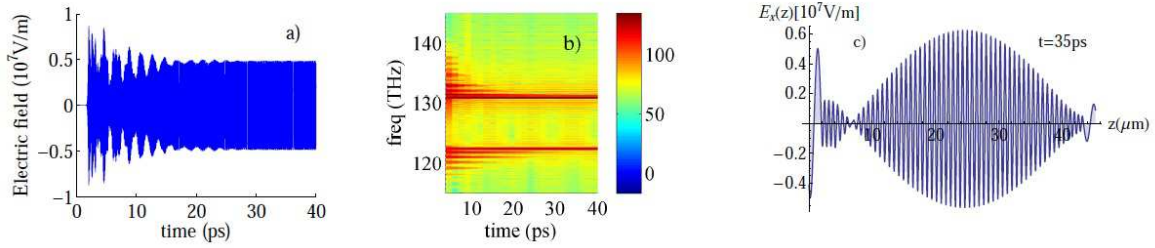}
\caption{a) Time-dependent output intensity in the left side of the BS chain; mode beating is observed in the time range $(1\div 30 ps)$ b) time evolution of its spectrum; c) snapshot of the electric field for t=35ps. \label{fig:f4}}
\end{figure*}
For the BS chain, with a gap in the range ($120\div 134$)THz, and $\phi=0.7\pi$, the TLS resonance frequency is $\nu_o=\nu_\ell (0.7\pi)$. 
Other parameters are: N=$10^{23}m^{-3}$, $\gamma=4.8 \times 10^{-28}Cm$, $T _1=1/\gamma_1=10^{-11}s$ and $T_2=1/\gamma_2=7*10^{-15}s$. 
For this system, Fig.~\ref{fig:f4} a) and b) show the time-dependent output intensity in the left side of the structure, and the time-resolved spectrum. After a wide-band transient (t$\approx$ 1ps), for $t \in \,(1 \div 30)\,ps$ emission is multimodal with a spectrum corresponding to delocalized Bloch modes at the PBG band-edges. At longer times (t $\gg$ 30 ps), the high quality factor modes survive and the spectrum is characterized by two main peaks: the one at shorter wavelengths corresponding to the PBG lower edge and the one inside the gap corresponding to the edge-state $\lambda_\ell (0.7\pi)$=2469 nm.
This is confirmed by the electric field spatial profile in Fig.~\ref{fig:f4}c) for $t=35$~ps, which reveals the coexistence of a extended mode and a localization at $z\cong4\mu$m.  

The US chain provides similar results. 
From eq.(4) and (1), for given $N$ and $\Gamma_o$, the dipole coupling coefficient is fixed by $N\gamma^2=w \Gamma_o \chi_A(\omega_o)$. 
On the other hand the $\Gamma_o$ value allows to control the gap width $\Delta\omega=2w_U\Gamma_o$ and the lefte-dge mode resonance $\omega_{\ell}=w_L\Gamma_o+\omega_o$. This circumstance allows a tunable field emission, either varying $\Gamma_0$ or the pumping rate $N$.
We choose $\xi=0.5\pi$, furnishing $w=w_U=0.4179$ and $w=w_L=-0.3465$ for the stop band upper edge and left-edge mode in Fig.~\ref{fig:f12}b).    

\begin{figure}[h!]
\begin{minipage}{0.52\columnwidth}
\includegraphics[width=1\columnwidth]{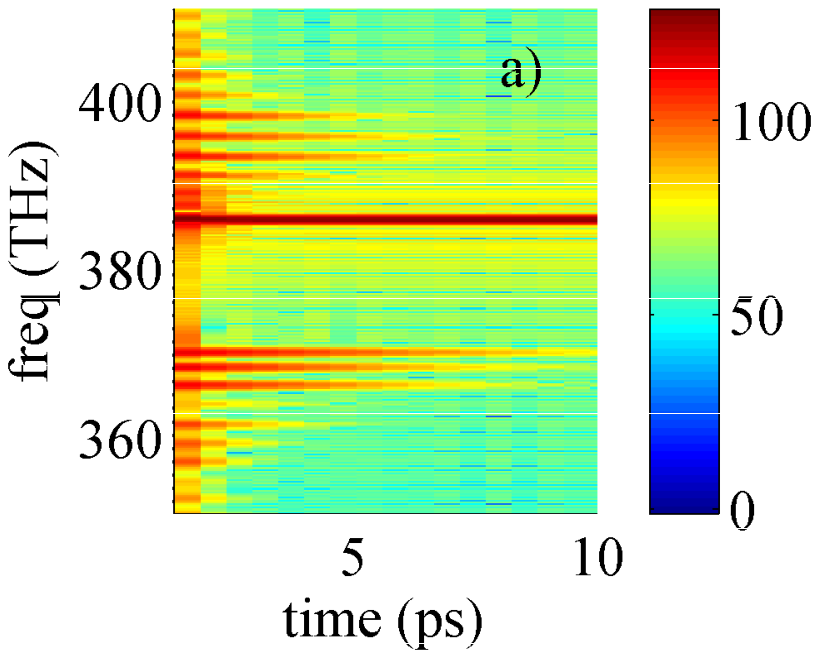}
\end{minipage}
\begin{minipage}{0.46\columnwidth}
\includegraphics[width=0.7\columnwidth]{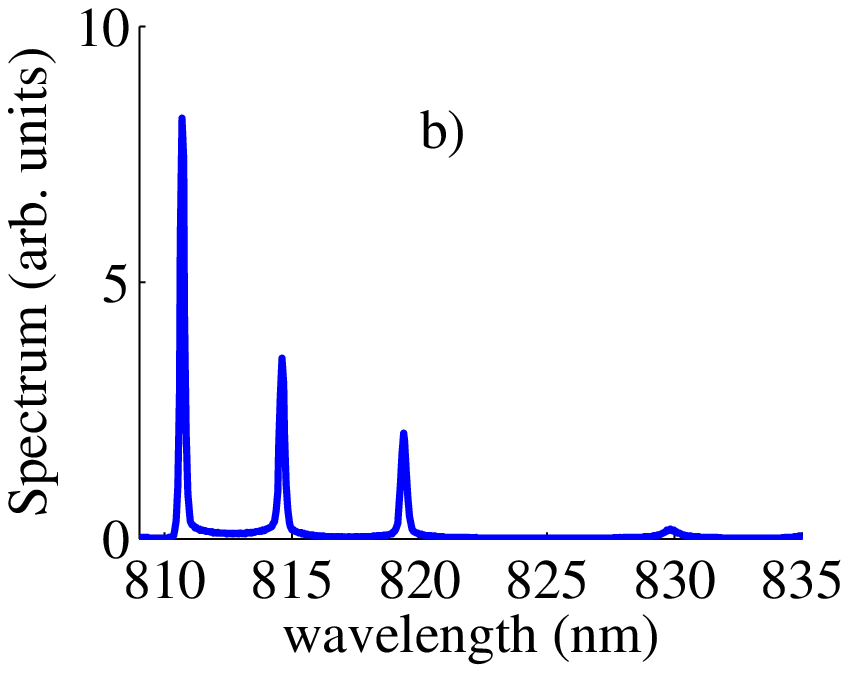}
\\[3mm]
\includegraphics[width=0.7\columnwidth]{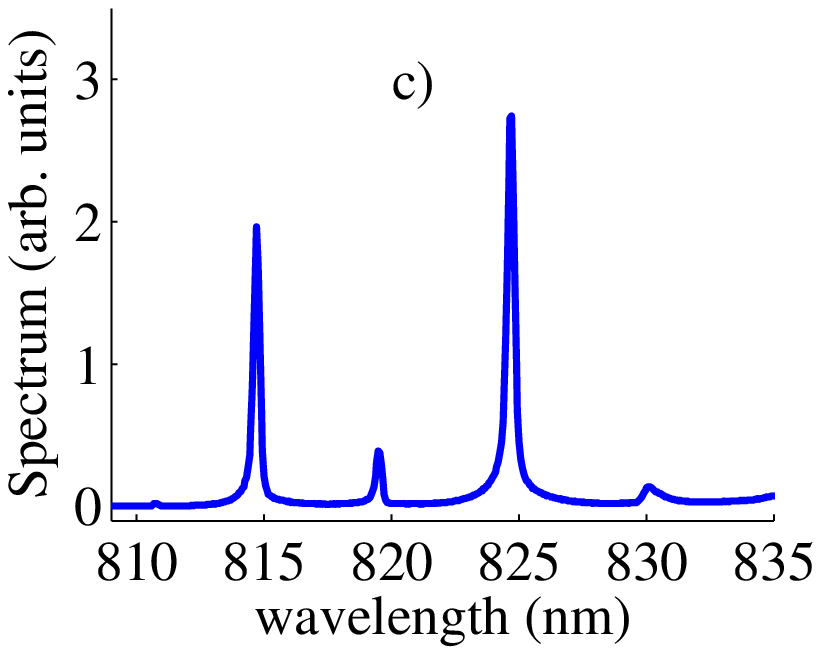}
\end{minipage}
\caption{a) Time-dependent spectrum of the output signal in the left side of the US chain. Emitted spectrum for b) $\Gamma_o=5.89*10^{-3}eV$ and c) $\Gamma_o=2.945*10^{-2}eV$ \label{fig:f5}}
\end{figure}

As shown in Fig.~\ref{fig:f5}a) for N=$10^{23}m^{-3}$ and $\Gamma_o=5.89*10^{-3}eV$, stimulated emission starts to overtake the spontaneous one after a transient regime (t$\approx$ 3ps) of laser field build-up. Once the steady state is reached the spectrogram of the emitted signal shows characteristic peaks. The peak at $\nu\approx$ 385 THz corresponds to a delocalized mode. In addition, the optical feedback 
edge-mode localization gives rise to emission at $\nu_\ell$=370.2 THz. For this system a wavelenghts tuning of the emitted spectrum can be obtained changing the $\Gamma_o$ value as shown in Fig.~\ref{fig:f5}b) for $\Gamma_o=5.89*10^{-3}eV$ where $\lambda_\ell$=810.2 nm and ~\ref{fig:f5}c) $\Gamma_o=2.945*10^{-2}eV$ where $\lambda_\ell$=814 nm.

\textit{Conclusions ---} In this paper we have analyzed the time-resolved optical response to an ultrashort light pulse and focused on edge states detection in 1D resonant topological insulators given by two-level layers in uniform and modulated refractive index structures.
For favorable system parameters obtained trough linearized Maxwell-Bloch equations, we show that a direct observation of topological protected edge states can be achieved following the time evolution of the population 
inversion with different properties of uniform structures with respect to periodic systems.
We provide evidence that a RTI can act as a resonator with laser like emissions due to localized edge modes;
we also show that the emission frequency can be tuned by acting on the pumping energy or other system parameters.

An experimental test of our results is possible by the use of  active TLS of quantum wells embedded in a semiconductor structure with periodically alternating linear index of refraction. In fact, for low densities, excitons in quantum wells can be considered as effective two-level systems if their resonance is close to the operating frequency. For these systems, the mechanism of emission is expected to have a low or vanishing laser threshold since, beeing the resonator directly etched in the amplifying material, an effecting feedback can be obtained.

Using topologically protected states for lasing in resonant systems may open a variety of several new directions in laser physics. Achieving lasing-like action may be favored in regimes in which no feasible way for invertion population can be imagined as for example silicon lasers; in addition topologically protected states may also allow to have very narrow band emission because of the low coupling with radiation modes, proving extremely coherent sources at room temperature for metrological and spectroscopic applications.

\begin{acknowledgments}
We acknowledge support from the ERC project VANGUARD (grant number 664782), and the Templeton Foundation (grant number 58277).
\end{acknowledgments}

\end{document}